\documentclass[aps,showpacs,superscriptadress,preprint]{revtex4}

\usepackage{graphicx}

\usepackage{dcolumn}

\usepackage{bm}

\usepackage{here}

\usepackage{amsmath}

\begin{document}

\title{The generalized second law of thermodynamics in the accelerating universe}

\author{Jia Zhou, Bin Wang}
\email{wangb@fudan.edu.cn} \affiliation{Department of Physics,
Fudan University, 200433 Shanghai}

\author{Yungui Gong}
\email{yungui_gong@baylor.edu} \affiliation{School of Physical
Science and Technology, Southwest University, Chongqing 400715,
China} \affiliation{CASPER, Department of Physics, Baylor
University, Waco, TX 76798, USA}

\author{Elcio Abdalla}
\email{eabdalla@fma.if.usp.br} \affiliation{Instituto de Fisica,
Universidade de Sao Paulo, C.P.66.318, CEP 05315-970, Sao Paulo}

\begin{abstract}

We show that in the accelerating universe the generalized second law
of thermodynamics holds only in the case where the enveloping
surface is the apparent horizon, but not in the case of the event
horizon. The present analysis relies on the most recent SNe Ia
events, being model independent. Our study might suggest that event
horizon is not a physical boundary from the point of view of
thermodynamics.

\end{abstract}


\maketitle

Today there is an overwhelming amount of data showing that our
universe is experiencing an accelerated expansion driven by a so
called "dark energy". The nature of such previously unforeseen
energy still remains a complete mystery, except for the fact that it
has negative pressure. In this new conceptual set up, one of the
important questions concerns the thermodynamical behavior of the
accelerated expanding universe driven by dark energy. It is
interesting to ask whether thermodynamics in the accelerating
universe can tell us some properties of dark energy. There is at
least a possibility to answer this question: it has been argued that
there is a deep connection between thermodynamics and gravity after
the discovery of the relation between thermodynamics and gravity in
black hole physics in 1970's \cite{Beken}. Thoughts on the profound
relation between thermodynamics and gravity have been inspired by
general situations, including the cosmological context \cite{Jacob}.
A close connection was disclosed between thermodynamics and
Friedmann equation derived in Einstein gravity in the cosmological
cases \cite{Cai}. This connection implies that the thermodynamical
properties can help understand the dark energy, which gives strong
motivation to study thermodynamics in the accelerating universe.

The thermodynamics of the de Sitter universe was considered some
years ago \cite{Hawking}. The de Sitter universe experiences
accelerated expansion and has only one cosmological horizon
analogous to the black hole horizon. It was found that the first law
of thermodynamics is valid at this horizon. If the accelerating
universe is driven by dark energy with equation of state $w\neq -1$,
the single horizon in the de Sitter case will be separated into the
apparent horizon and the event horizon. The apparent Hubble horizon
of the universe always exists while the event horizon does not. The
thermodynamical properties associated with the apparent horizon have
been found in a quasi-de Sitter geometry of inflationary universe
\cite{22}. In the late time accelerating universe, it was disclosed
that the first law of thermodynamics holds in the physically
relevant part of the universe enveloped by the dynamical apparent
horizon, while does not hold appropriately in the region enveloped
by the cosmological event horizon\cite{wang}. This result is
supported if one considers the relation between thermodynamics and
gravity \cite{Cai}.

Besides the first law of thermodynamics, a lot of attention has been
paid to the generalized second law of thermodynamics in the
accelerating universe driven by dark energy
\cite{Pavon,wang2,gong2}. The generalized second law of
thermodynamics is as important as the first law, governing the
development of the nature. Using a specific model of dark energy,
the generalized second law as defined in the region enveloped by the
apparent horizon as well as in the event horizon was examined in
\cite{wang}. It was found that it is obeyed in the case of the
universe enveloped by the apparent horizon, not otherwise.

In this paper we shall extend our study on the generalized second
law of thermodynamics to a general accelerating universe with model
independent dark energy equation of state. We use the most recent
type Ia supernovae observations to carry out the model independent
analysis of the generalized second law of thermodynamics. Recently,
the SNe Ia data have been used to examine the energy conditions'
violation in the context of the standard cosmology
\cite{gong3a,gong3b}.

Within the framework of the standard FRW cosmology,
\begin{equation}
ds^{2}=-dt^{2}+a^{2}(t)(dr^{2}+r^{2}d\sigma^{2})\quad ,
\end{equation}
the evolution of the universe is governed by Friedmann equations
\begin{eqnarray}
H^{2}&=&\frac{8\pi}{3}\rho\quad ,\\
\dot{H}&=&-4\pi(\rho+P)\quad ,
\end{eqnarray}
where $H=\dot{a}/a$ is the Hubble parameter. Assuming a perfect
cosmological fluid, we have
\begin{equation}
\dot{\rho}+3H(\rho+P)=0\quad ,
\end{equation}
where $\rho, P$ are respectively the energy density and the pressure
of the total content of the universe.

The entropy of the universe inside the horizon can be related to
its energy and pressure in the horizon by the Gibbs equation
\cite{Pavon}
\begin{equation}
TdS_{in}=d({\rho}V)+PdV=Vd{\rho}+(\rho+P)dV.
\end{equation}
For the dynamical apparent horizon $R_A=1/H$. Considering the
total volume $V=\frac{4\pi}{3}R^{3}_{A}$, we have
\begin{equation}
TdS_{in}=(H^{-2}+\dot{H}H^{-4})dH.
\end{equation}
We assume that the temperature scales as the temperature of the
horizon, which we assume to be $T_H=H/(2\pi)$. It is natural to
suppose that the temperature of the perfect fluid inside the
apparent horizon is $T=bT_H$, where $b$ is a real proportional
constant to be figured out\cite{Pavon,Sad}. We limit ourselves to
the assumption of the local equilibrium hypothesis that the energy
would not spontaneously flow between the horizon and the fluid,
the latter would be at variance with the FRW geometry.

In addition to the entropy of the universe inside the apparent
horizon, there is a horizon entropy associated to the apparent
horizon,
\begin{equation}
S_{A}={\pi}R_{A}^{2}=\pi{H^{-2}}\quad .
\end{equation}

In order to check the generalized second law of thermodynamics, we
have to examine the evolution of the total entropy $S_{in}+S_A$,
which is in the form
\begin{equation}
\dot{S}_{in}+\dot{S}_{A}=2\pi{H^{-5}\dot{H}[(1-b)H^{2}+\dot{H}]b^{-1}}.
\end{equation}

This expression holds in the local equilibrium hypothesis\cite{de
Groot}. If the temperature of the fluid differs much from that of
the horizon, there will be spontaneous flow between the horizon
and the fluid and the local equilibrium hypothesis will no longer
hold. For the non-equilibrium case, an interaction term should be
added on the right-hand-side of the above equation which will make
the discussion more complicated\cite{Jou}. Here we limit our
discussion on the local equilibrium assumption.

For $b\leq 1$, the generalized second law can be secured provided
that
\begin{equation}
\dot{H}\leq{(b-1)H^{2}} \hspace{0.5cm} {\rm or} \hspace{0.5cm}
\dot{H}\geq{0}\quad . \label{9}
\end{equation}
However, in the range $(b-1)H^2<\dot{H}<0$, the generalized second
law breaks down. The first integral of the above equation leads to
\begin{equation}
\dot{a}\geq{H_{0}a_{0}^{-b+1}a^{b}}\hspace{0.5cm} {\rm or}
\hspace{0.5cm}  \dot{a}\leq{H_{0}a}\quad .\label{10}
\end{equation}

Using definitions for the radial coordinate distance, the luminosity
distance and distance modulus,
\begin{eqnarray}
r(a)&=&\int_{0}^{r(a)}dr=\int_{a}^{a_{0}}\frac{da'}{\dot{a}'a'}
=\int_{a}^{a_{0}}\frac{da'}{Ha'^{2}}\quad ,\label{11}\\
d_{L}(z)&=&a_{0}(1+z)r(a)\quad ,\label{12}\\
\mu(z):&=&m(z)-M=5\lg{d_{L}(z)}+25\quad ,\label{13}
\end{eqnarray}
we obtain the upper or lower bound on the radial distance necessary
to enforce the generalized second law as
\begin{equation}
r(z)\leq{\frac{(1+z)^b-1}{H_{0}a_{0}b}} \hspace{0.5cm} {\rm or}
\hspace{0.5cm} r(z)\geq{\frac{z}{H_{0}a_{0}}}\quad ,\label{14}
\end{equation}
where $a_0/a=1+z$. Combining (10), (12), (13) and (14), we obtain
inequalities for the distance modulus
\begin{equation}
\mu(z)\leq{5\lg{\frac{(1+z)[(1+z)^b-1]}{H_{0}b}}+25}
\hspace{0.5cm} {\rm or} \hspace{0.5cm}
\mu(z)\geq{5\lg{\frac{(1+z)z}{H_{0}}}+25}\quad .\label{15}
\end{equation}
The bounds on the distance modulus as a function of the redshift are
shown in Fig.1a, where in the shadow the generalized second law
breaks down.

For $b\geq1$, the generalized second law holds if
\begin{equation}
\dot{H}\geq(b-1)H^{2}\hspace{0.5cm} {\rm or} \hspace{0.5cm}
\dot{H}\leq 0\quad .\label{16}
\end{equation}
From these constraints we find that the generalized second law
cannot hold when $0<\dot{H}<(b-1)H^{2}$. The bounds for enforcing
the generalized second law on the distance modulus are
\begin{equation}
\mu(z)\geq{5\lg{\frac{(1+z)[(1+z)^b-1]}{H_{0}b}}+25}
\hspace{0.5cm} {\rm or} \hspace{0.5cm}
\mu(z)\leq{5\lg{\frac{(1+z)z}{H_{0}}}+25}\quad , \label{17}
\end{equation}
which are shown in Fig.1b. The area in shadow violates the
generalized second law.
\begin{figure}[tbp]
\includegraphics[width=7.8cm,height=7.0cm]{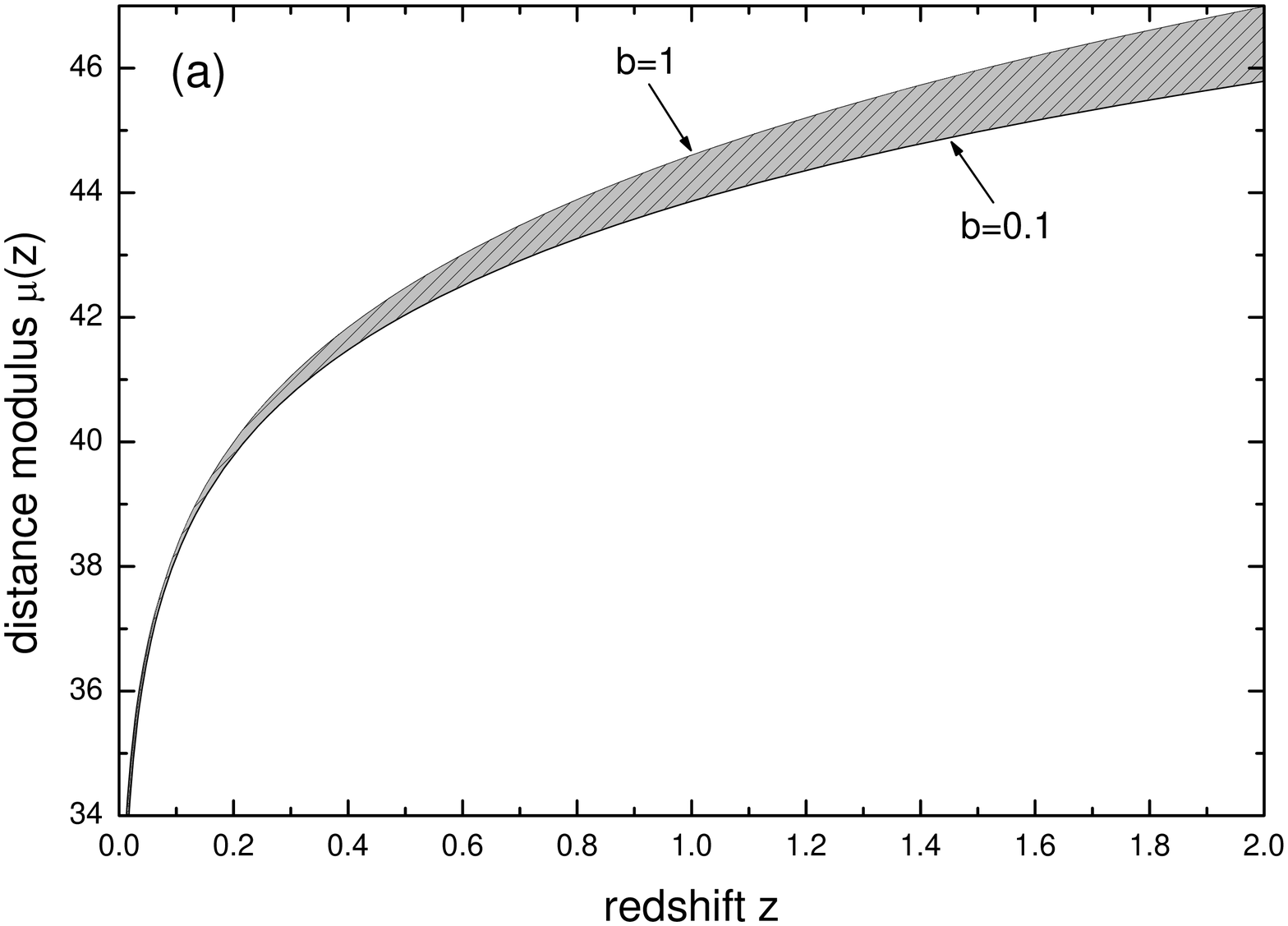}
\includegraphics[width=8.5cm,height=7.2cm]{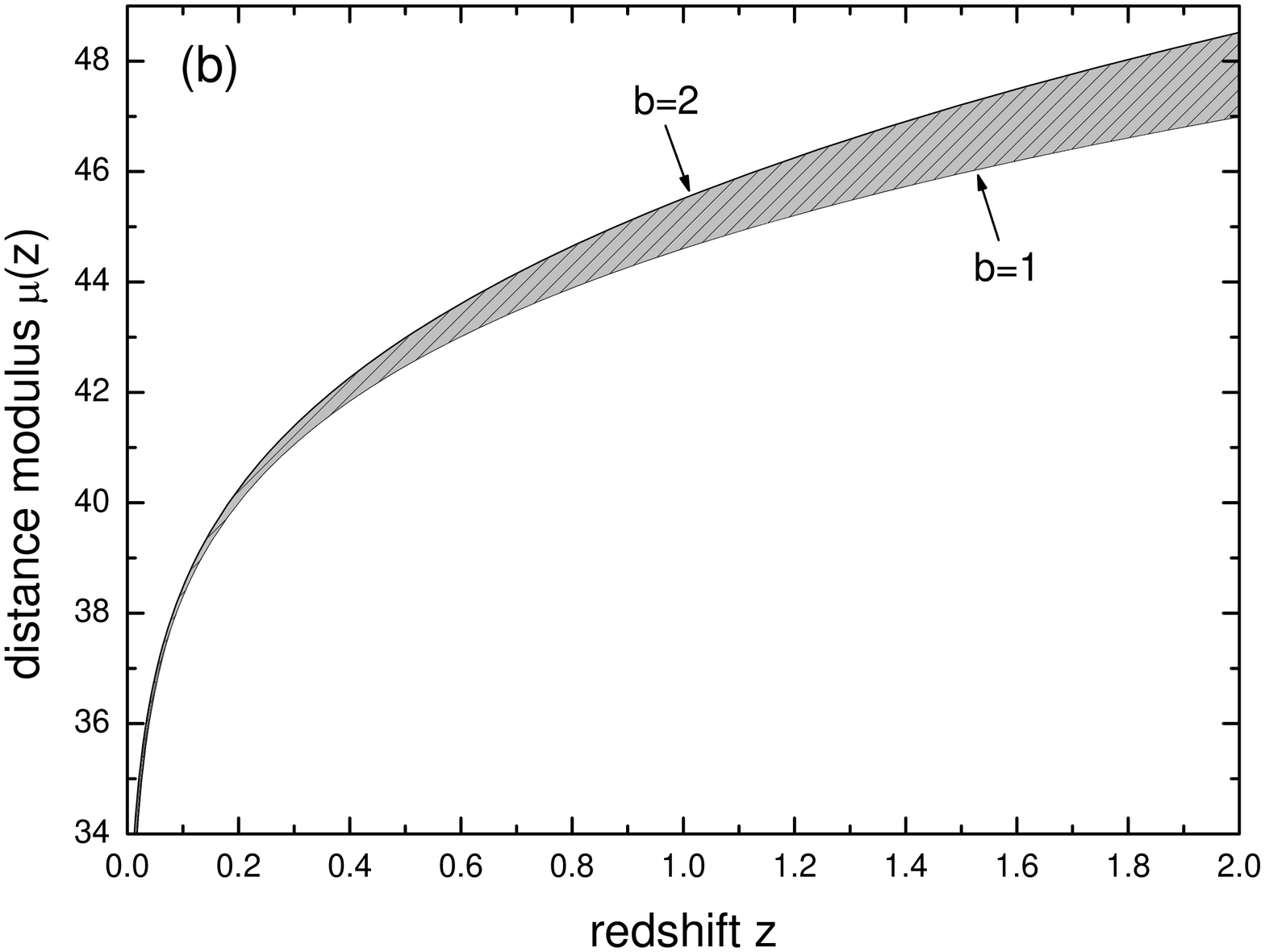}
\caption{(a) shows the condition to protect the generalized second
law for $b\leq 1$ and (b) shows the condition for $b\geq 1$. We use
the central value for the Hubble parameter,
$H_{0}=72km{\cdot}s^{-1}{\cdot}Mpc^{-1}$ \cite{Freedman}. }
\end{figure}

The shadows in Fig.1 shrink when $b\rightarrow 1$ and disappear when
$b=1$. The generalized second law of thermodynamics is one of the
most general laws governing the nature. Requiring validity of the
generalized second law leads to the limit $b\rightarrow 1$. This
requires that the temperature of the fluid inside the universe
should be in equilibrium with the apparent horizon temperature. The
fluid and the horizon must interact for some length of time but
finally they must attain thermal equilibrium\cite{Pavon}. In the
thermal equilibrium, the generalized second law holds for any value
of $\dot{H}$.

When the universe is in equilibrium with the apparent horizon
temperature, the line for the distance modulus corresponds to
$\dot{H}=0$, which shows that the total entropy of the universe
keeps unchanged. If the observational data accumulate on this
line, we will learn that the universe is the de Sitter space with
$w=-1$.

In order to shed some light on the generalized second law in the
universe from the observational side, it is important to confront
the condition of protecting the generalized second law with the
current observational data. In this regard, we use the most recent
SNe Ia gold sample compiled in\cite{Riese}. This sample consists of
182 data, in which 16 points with $0.46<z<1.39$ were discovered by
HST, 47 points with $0.25<z<0.96$ from the first year SNLS and the
left 119 old data. Results are shown in Fig.2 where we take $b=1$.
Panels (2a) and (2b) show the condition for the generalized second
law and data points in the small redshift $z<0.1$ and $0.1\leq z\leq
0.27$ respectively, while panel (2c) shows the curve and data for
redshift interval $0.27\leq z\leq 2$.
\begin{figure}[tbp]
\includegraphics[width=8.0cm,height=7.0cm]{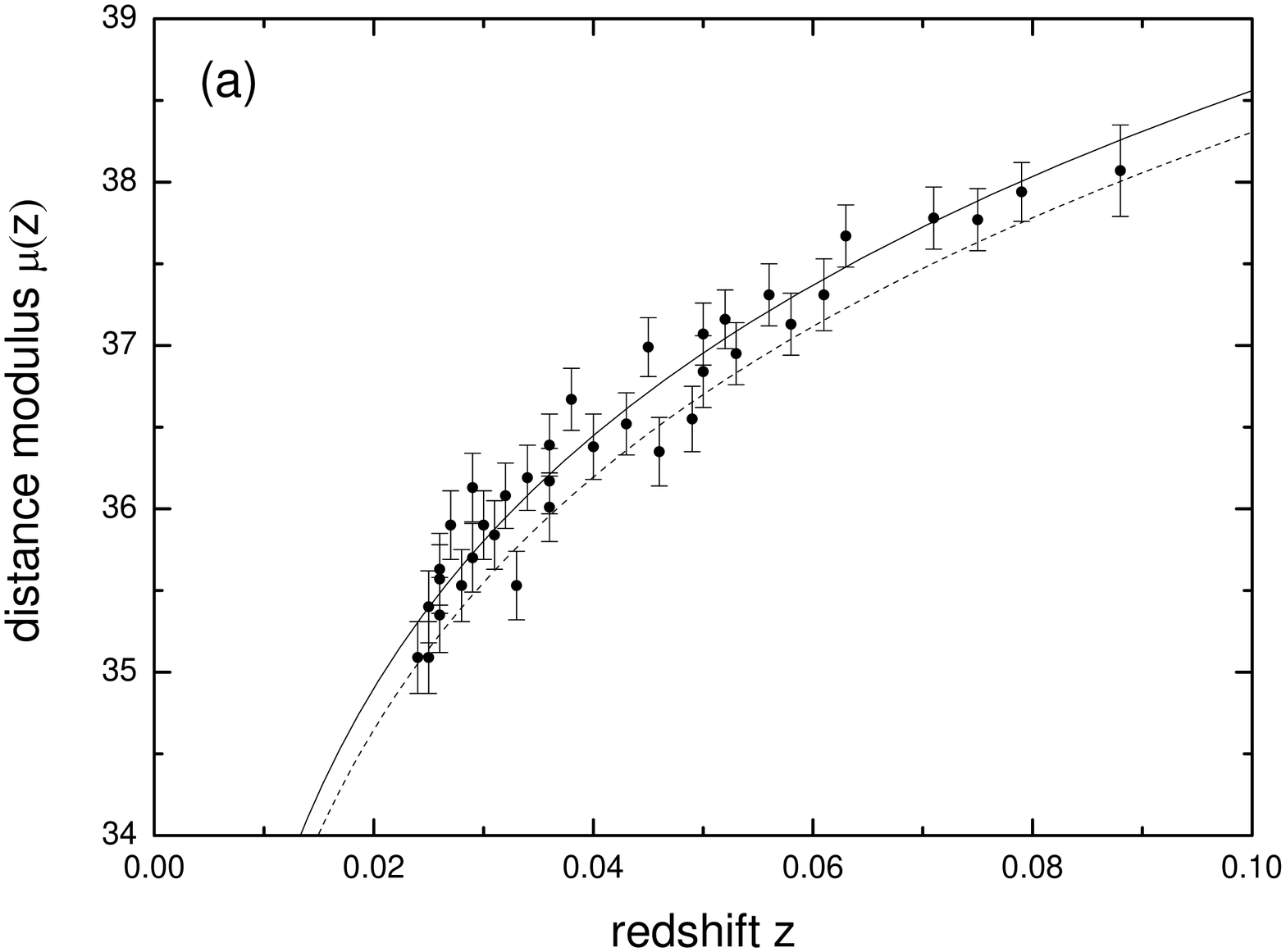}
\includegraphics[width=8.0cm,height=7.0cm]{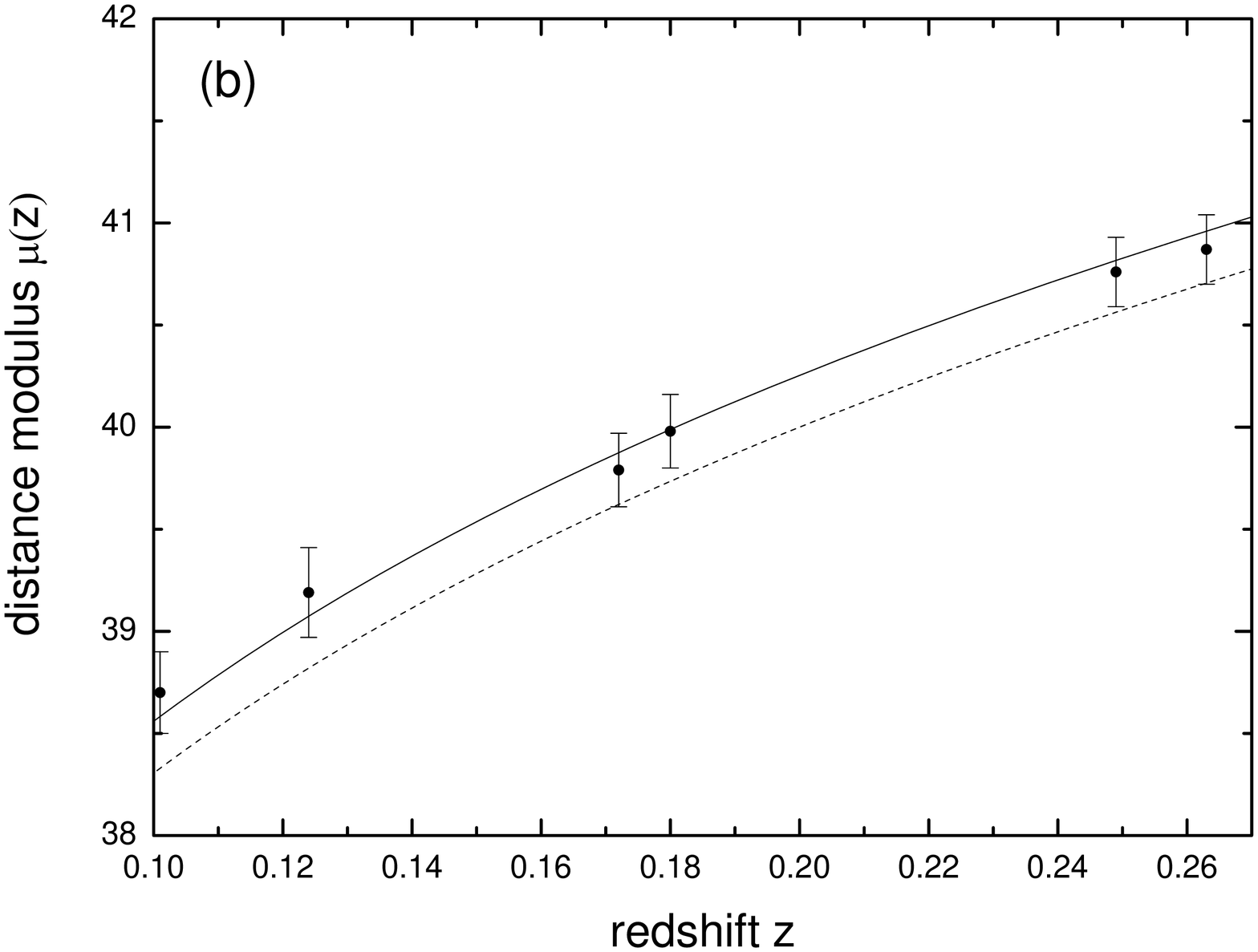}
\includegraphics[width=8.0cm,height=7.0cm]{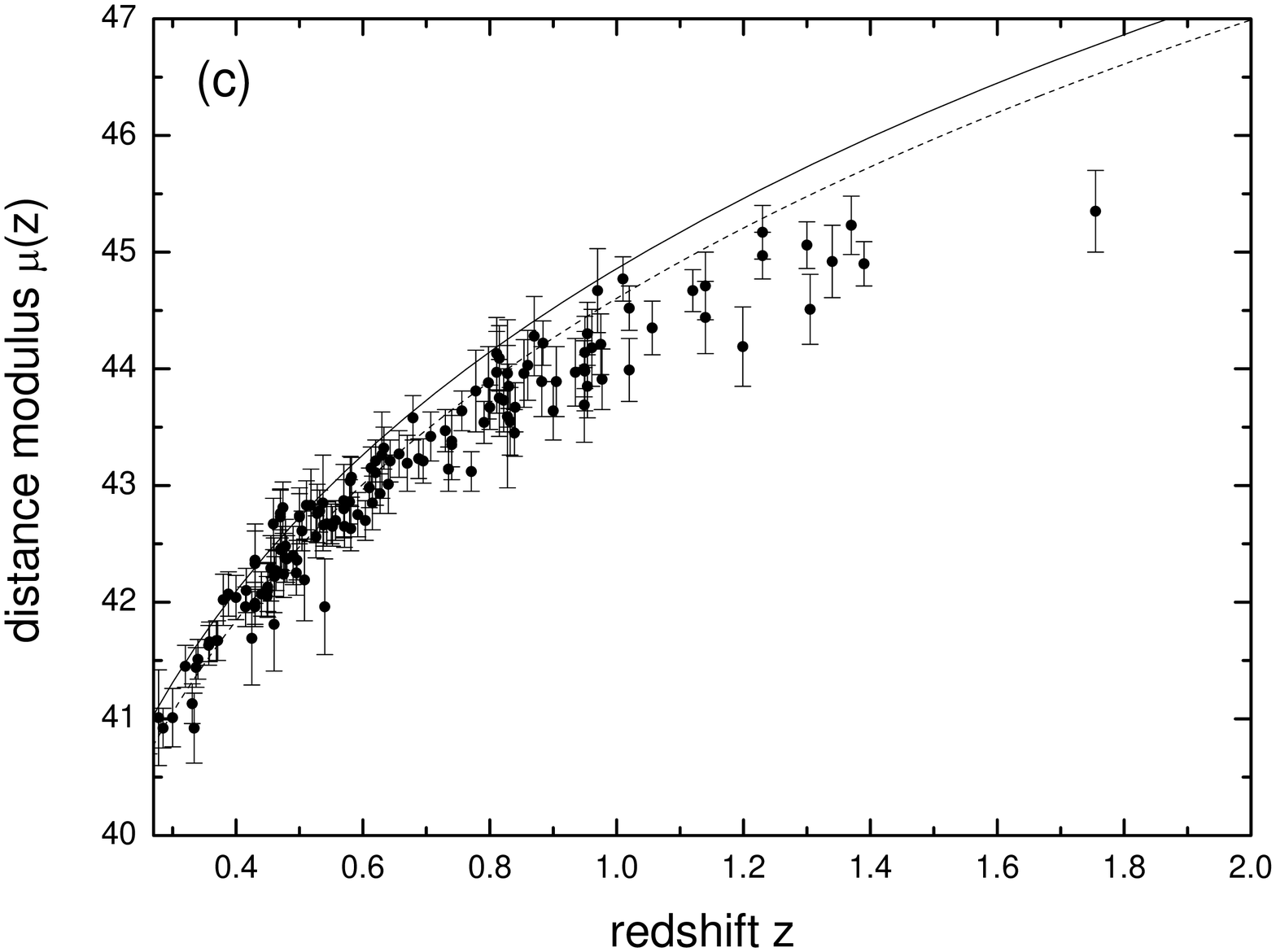}
\caption{Panels (2a) and (2b) show the condition for the generalized
second law and data points in the small redshift $z<0.1$ and
$0.1\leq z\leq 0.27$ respectively, while panel (2c) shows the curve
and data for redshift interval $0.27\leq z\leq 2$. The dashed line
was plotted by using $H_{0}=72km{\cdot}s^{-1}{\cdot}Mpc^{-1}$, while
the solid line is for $H_{0}=64.04km{\cdot}s^{-1}{\cdot}Mpc^{-1}$. }
\end{figure}

The dashed line was plotted by choosing $H_0=72Kms^{-1}Mpc^{-1}$,
which is the central value of different observations for the present
Hubble parameter and this value was also taken in plotting the
distance modulus in \cite{gong3a}.  It is interesting to see that
the data follow the line, but are not exactly on the line. This
tells us that $\dot{H}$ is not always zero so that our universe is
not exactly de Sitter's. For the small redshift $z\leq 0.27$, the
majority of SNe events are above the dashed line, while for the
bigger redshift $0.27\leq z\leq 2$, the majority of the SNe events
are below the dashed line. For the low redshift $(z\leq 0.27)$, the
SNe events above the dashed line do not exactly lead $\dot{H}>0$.
This is because we cannot exactly obtain (\ref{9}) from (\ref{15})
or (\ref{16}) from (\ref{17}). The determination of when $\dot{H}>0$
is important, since this can tell us the moment when the universe
starts to super-accelerate. However the integrated over results in
(15)(17) hide the information of $H$. To make clear the behavior of
$\dot{H}$ for the SNe data in the low redshift, we can define
$f(z)=r(z)-\frac{z}{H_{0}a_{0}}$, where $\frac{z}{H_{0}a_{0}}$ is
the value of $r(z)$ when $\dot{H}=0$. Its first derivative with
respect to $z$ reads $f'(z)=\frac{dr}{dz}-\frac{1}{H_{0}a_{0}}$. At
$z=0, f(0)=0$ and $f'(0)=0$. In the vicinity of $z=0, f(0^+)>0$, so
that $f'(0^{+})=\lim_{\varepsilon\rightarrow0^{+}}
\frac{f(\varepsilon)-f(0)}{\varepsilon}>0$ and
$f''(0^{+})=\lim_{\varepsilon\rightarrow0^{+}}
\frac{f'(\varepsilon)-f'(0)}{\varepsilon}>0$. The second derivative
of $f(z)$ with respect to $z$ has the form
$f''(z)=\frac{d^{2}r}{dz^{2}}=\frac{a\dot{H}}{a_{0}H^{3}}$. It's
easy to see that $f''(z)>0$ corresponds to $\dot{H}>0$ and in this
period $f(z)$ experiences a fast increase with $z$. When $\dot{H}=0,
f'(z)$ reaches its maximum. For $\dot{H}<0$, $f(z)$ keeps positive
and slowly increases with $z$. The position where $f'(z)$ reaches
the maximum (for $\dot{H}=0$) can be read from the data since
$\frac{dr}{dz}=\frac{10^{\frac{\mu-25}{5}}}{a_{0}(1+z)}(\frac{\ln10}{5}\frac{d\mu}{dz}-\frac{1}{1+z})$.
Using the finite difference method, we show approximate values of
$\frac{\Delta\mu_0}{\Delta z}$ and $a_0\frac{\Delta r}{\Delta z}$ by
employing the central values of distance modulus of SNe events in
the table below. It is clear that there is a maximum value of
$\frac{\Delta\mu}{\Delta z}$ around $z_t\sim 0.101$, which
corresponds to the biggest jump of $\Delta r/\Delta z$ where $f'(z)$
reaches its maximum value. We learn that when $z<z_t, \dot{H}>0$,
while $\dot{H}<0$ for $z>z_t$. Thus the SNe data indicate that $z_t$
is a turning point indicating the transition to the
super-acceleration at this redshift. This is consistent with the
phantom divide position found in \cite{acc}. We expect more accurate
SNe data to make the transition position to be determined more
exactly. If just from the data listed below, it is also possible
that the universe evolves with the oscillating equation of state
below and above $-1$, since there are peaks of $\Delta\mu/\Delta z$
and $\Delta r/\Delta z$ around $z=0.101, 0.180$.

\vspace{0.6cm}
\begin{tabular}
{p{1.5cm}p{1.85cm}p{1.85cm}p{1.85cm}p{1.85cm}p{1.85cm}p{1.85cm}p{1.5cm}}
\multicolumn{8}{c}{Some observational data above the line $\dot{H}=0$} \\
\hline\hline $z$ &0.088 &0.101 &0.124 &0.172 &0.180 &0.249 &0.263\\
\hline
$\mu_{0}$ &38.07 &38.70 &39.19 &39.79 &39.98 &40.76 &40.87\\
\hline $\frac{\Delta\mu_{0}}{\Delta z}$ &34.55 &48.46 &21.30
&23.75 &23.75
&11.30 &7.86\\
\hline $a_{0}\frac{\Delta r}{\Delta z}$ &5665.31 &10685.55
&5464.70
&7811.00 &8472.30 &5002.72 &3342.42\\
\hline\hline
\end{tabular}
\vspace{0.6cm}

In plotting the distance modulus, it was argued that the zero
point is arbitrary and a proper value needs to be subtracted from
the distance modulus if we use
$H_0=72Kms^{-1}Mpc^{-1}$\cite{Riese}. The same effect to overcome
the arbitrariness can be achieved by getting the $H_0$ from the
nearby SNe data $z\leq 0.1$ using $d_L(z)=H_0 z$ and it was found
that $H_0=64.04Kms^{-1}Mpc^{-1}$\cite{gong3b}. Using
$H_0=64.04Kms^{-1}Mpc^{-1}$, in fig.2 we plotted the solid line
and we see that nearly the same amount of SNe events are above and
below the solid line for $z<0.1$. We see that the normalization
makes it even more difficult to disclose the super-acceleration
behavior of the universe from the integrated effect.

Recently Simon et al \cite{Simon} have published Hubble function
$H(z)$ data extracted from differential ages of passively evolving
galaxies. The Hubble function is not integrated over, which
contains some fine structures covered by the distance modulus. It
was observed that $H(z)$ decreases with respect to the redshift
$z$ around $z\sim 0.3$ and $z\sim 1.5$ \cite{wei}, which disclosed
the moment when the universe may enter super-acceleration with the
total fluid behaves like phantom. The oscillation feature of the
equation of state was also observed in studying the Hubble
parameter data.

Now we extend our discussion to the universe enveloped by the
cosmological event horizon with the volume
$V=\frac{4\pi}{3}R_{E}^{3}$, where $R_E$ is the event horizon,
defined by $R_{E}=a(t)\int_{t}^{t_{\max}}\frac{dt'}{a(t')}$. From
the Gibbs law, we find that the entropy inside the event horizon
is expressed as
\begin{equation}
TdS_{in}=\frac{4\pi}{3}R_{E}^{3}\dot{\rho}dt-R_{E}^{2}
\dot{H}(HR_{E}-1)dt=\dot{H}R_{E}^{2}dt\quad .\label{18}
\end{equation}
Including the event horizon entropy $S_{E}=2{\pi}R_{E}^{2}$ and
supposing $T=b/(2\pi R_E)$, the change of the total entropy has the
form
\begin{equation}
\dot{S}_{in}+\dot{S}_{E}=2{\pi}R_{E}(b^{-1}R_{E}^{2}
\dot{H}+\dot{R}_{E})\quad .\label{19}
\end{equation}
in the local equilibrium hypothesis. In order to keep the
generalized second law of thermodynamics, we require the above
equation to be non-negative, which leads
\begin{equation}
H_{0}R_{E}-HR_{E}\geq b\frac{R_{E}}{R_{E_{0}}}-b,
\end{equation}
where $R_{E_{0}}=a(t_{0})\int_{t_{0}}^{t_{\max}}\frac{dt}{da}$ is
the current value of the future event horizon. Using the
definition of the radial distance, we have
$R_{E}=ar+\frac{a}{a_{0}}R_{E_{0}}$, and the above equation can be
written as
\begin{equation}
H_{0}ar'+a^{-1}r'\frac{da}{dr'}\geq \frac{b}{R_{E_{0}}}ar'-b,
\end{equation}
where $r'=r+\frac{R_{E_{0}}}{a_{0}}$. Considering
$\frac{da}{dr'}=-\frac{a^{2}}{a_{0}}\frac{dz}{dr'}$ and taking
$x=a_{0}r'$, one gets
\begin{equation}
\frac{dz}{dx}-\frac{b(1+z)}{x}\leq{H_{0}-bR_{E_{0}}^{-1}}\quad .
\end{equation}
Simply choosing the proportional constant $b$ as unity which means
that the perfect fluid is in thermal equilibrium with the event
horizon, the generalized second law then requires
\begin{equation}
\frac{dz}{dx}-\frac{1+z}{x}=B, \hspace{0.5cm} B\leq
H_{0}-R_{E_{0}}^{-1}\quad .
\end{equation}
Thus, we have
\begin{equation}
z=Ax+Bx{\ln}x-1\quad .
\end{equation}
Taking account of the present values $z_{0}=0, x_{0}=R_{E_{0}}$, $
A=x_{0}^{-1}-B{\ln}x_{0}$ we have
\begin{equation}
z=y+Bx_{0}y{\ln}y-1\quad ,
\end{equation}
where $y=x/x_0=(a_0 r+R_{E0})/R_{E0}$. This equation shows the
relation between the redshift and the radial distance, from which we
can learn the behavior of the distance modulus for different $z$.
The generalized second law of thermodynamics requires $Bx_0\leq
H_{0}x_0-1$. Therefore the generalized second law is satisfied in
the region above the line of the distance modulus with $Bx_0=H_0 x_0
-1$.
\begin{figure}[tbp]
\includegraphics[width=8cm,height=7.0cm]{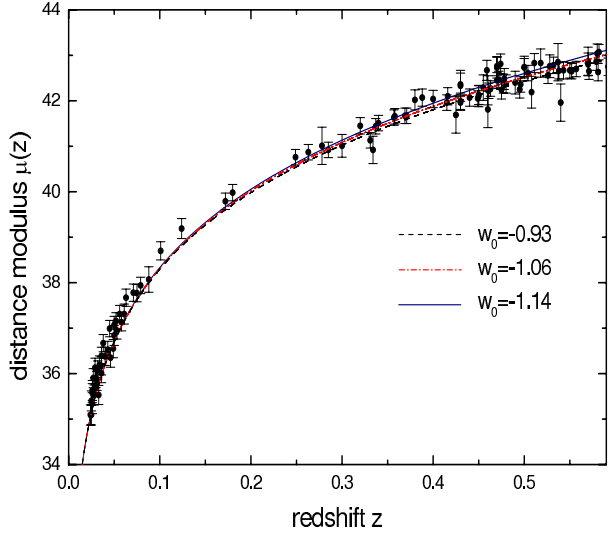}
\caption{The distance modulus with $Bx_0=H_0 x_0 -1$ when the
equation of state $w$ is taken to be $-0.93$, $-1.06$, and $-1.14$
respectively. New SNe gold sample data are included. It indicates
that the condition for the generalized second law seems to violate
by a considerable number of nearby SNe Ia.}
\end{figure}

For the constant equation of state, defining $\epsilon=3(1+w)/2$,
from the Friedmann equation we have $a(t)=t^{1/\epsilon}$ and
$R_E=-\epsilon t/(\epsilon -1)$ and $R_A=\epsilon t$
\cite{wang,buosso}. For $0<\epsilon<1$, $-1<w<-1/3$, the universe is
in the accelerating quintessence-like space, while for $\epsilon<0$,
$w<-1$, the universe is in phantom phase. The apparent horizon and
the event horizon do not differ much, they are related by
$R_A/R_E=1-\epsilon$. Thus $H_0
x_0-1=R_{E0}/R_{A0}-1=-3(1+w_0)/(1+3w_0)$. For different values of
the present equation of state $w_0$, in Fig.3, we plotted the
distance modulus as a function of the redshift by taking
$H_0=72Kms^{-1}Mpc^{-1}$. Within the observational range of the
current equation of state $w_0=-1.06^{+0.13}_{-0.08}$\cite{wmap},
most of the data lie below the lines, which are in violation with
the generalized second law requirement. For more negative equation
of state or choosing the appropriate normalization of the zero point
in the distance modulus, more SN data will be in the region in
violation with the generalized second law. This problem cannot be
overcome even when we set $w_0=-0.34$. The observation tells us that
the universe enveloped by the event horizon cannot satisfy the
generalized second law. This result is general, which supports
previous study with the interacting dark energy model\cite{wang}.

In summary, by extending previous results for the interacting
holographic dark energy model \cite{wang} to the general dark energy
cases, we have examined the generalized second law of thermodynamics
in the accelerating universe. We have also confronted the
requirement of the generalized second law with the new gold sample
182 SNe observed events. On general grounds, our analysis indicates
that the apparent horizon is a good thermal boundary. The
accelerating universe enveloped by the apparent horizon satisfies
the generalized second law. The SNe events indicate that there is a
transition of the equation of state at low redshift, much clearer
result could be obtained from the more precise Hubble function data
in the future which is not integrated over . However the
accelerating universe inside the event horizon does not satisfy the
generalized second law. The event horizon in the accelerating
universe might not be a physical boundary from the thermodynamical
point of view. This result is supported by the new SNe Ia gold
samples.

So far our discussion is limited in the local equilibrium
hypothesis. Further consideration is called for provided that the
horizon and fluid are not in a thermal equilibrium situation.

\begin{acknowledgments}
This work was partially supported by  NNSF of China, Ministry of
Education of China and Shanghai Education Commission. E. Abdalla's
work was partially supported by FAPESP and CNPq, Brazil. Y.G. Gong
was supported by Baylor University, NNSFC under grant No.s
10447008 and 10605042, and SRF for ROCS, State Education Ministry.
\end{acknowledgments}

\end{document}